# Radiation and Heating of Rotating Neutral Particle in Close Vicinity to the Surface of Transparent Dielectric

G. V. Dedkov and A. A. Kyasov

Kabardino-Balkarian State University, Nanoscale Physics Group, Nalchik, Russia

We have calculated the intensity of electromagnetic radiation from particles rotating in close vicinity to a transparent dielectric plate. The radiation is a result of quantum and thermal friction occurring during particle rotation, while the driving factor of radiation is a decrease in the kinetic energy of particle. The characteristic frequency of radiation is determined by the angular velocity and the temperature of the plate and surrounding vacuum background. These results are challenging for experimental verification in future.

**1. Introduction**

Quantum and non-contact friction (Casimir friction) of two uncharged polarizable bodies moving relative to each other is a manifestation of quantum and thermal fluctuations of the electromagnetic field. Despite its minuteness, the effects associated with quantum and non-contact friction of bodies at finite temperatures attract growing interest of many researchers [1-19] (note that this list is far from being complete). Many features of these and related effects were successfully described within the framework of the fluctuation electromagnetic theory, assuming local equilibrium temperatures for the electromagnetic field and the bodies in relative motion, and relativistic formulation of the fluctuation-dissipation relations [1–3, 5, 6, 8–11, 13, 16–19]. Several other approaches have also been used [4, 12, 15, 20, 21]. The reality of Casimir friction and its experimental investigation were discussed in [22].

As was first noted in [2], another important consequence of the vacuum friction is associated with accompanying radiation. This issue has been investigated further in recent papers [20, 21, 23–27]. Particularly, the emission from a small particle in rotational [20, 21, 27] or translational motion in radiation background [23, 26] was studied, as well as the Cherenkov radiation from a particle moving parallel to a transparent dielectric [24], and between the two transparent dielectric plates in relative motion [25] .

In the present work, the radiation from a particle rotating in close vicinity to the surface of transparent dielectric is investigated. The presence of dense medium increases the fluctuation-electromagnetic interactions, such as for example, the friction torque acting on rotating particle in an evanescent electromagnetic field of the surface [17, 18]. Therefore, one can expect that radiation in this case is more intensive. We also plan to clarify the issue of particle heating and its connection to frictional torque and radiated power.

The configuration of spinning particle and dielectric surface is more advantageous as compared to the configuration where the particle is moving parallel to a dielectric surface, since the condition of Cherenkov's radiation in the latter case is very rigid: it is satisfied when the speed of particle relative to the surface exceeds the velocity of light in the medium: $V > c/n$. Thus, the velocity $V$ should be very high even in the case of moderate refraction index $n$. It is this fact and a very short time of interaction of particle with surface make it difficult experimental measurements of such radiation from neutral nanoparticles in the nearest future.

In the case of particle rotating with angular velocity $\Omega$ close to the surface one obtains the shifted frequencies $\omega \pm \Omega$ instead of the Doppler shifts in the frequency arguments of electromagnetic fields and dipole moments (products of the particle velocity vector and the photon wave vector in the case of translational motion). Due to this, we have no rigid restrictions on the value of $\Omega$. Rapidly rotating particles have already been obtained and controlled in electromagnetic traps [28]. Thus, the torques created by circularly polarized laser light have induced rotational motion of levitating graphene flakes of 1 $\mu m$ size at frequencies exceeding 1 MHz, and it is expected that this frequency can be increased up to 1 GHz.

**2. General equations and the rates of emission and absorption**

Figure 1 shows the geometric configuration used in what follows and the reference frames $\Sigma$ and $\Sigma'$ corresponding to a thick transparent dielectric plate and a particle. We treat a spherical particle of radius $R$ rotating with angular velocity $\Omega$ around the $Z$ axis directed normal to the surface of the plate, assuming that the particle is located at distance $z_0$ from the surface. For generality, we assume that the particle has a local temperature $T_1$, while a dielectric plate and vacuum background have the temperature $T_2$. Moreover, we assume that the particle is characterized by complex electric and magnetic polarizabilities $\alpha_e(\omega)$, $\alpha_m(\omega)$, and dielectric plate – by real refractive index $n$. The particle can be treated as a point-like fluctuating electric and magnetic moment **d** and **m** provided that

$$R \ll z_0, \ \Omega R/c \ll 1, \ R \ll \min(2\pi\hbar c/k_B T_{1,2}) \tag{1}$$

As one can see, conditions (1) can be satisfied in a wide range of distances $z_0$ and temperatures $T_{1,2}$ for nano- and microsized particles and realistic angular velocities $\Omega$.

In earlier papers [18] we calculated the heating rate and frictional torque of a particle rotating in close vicinity to a heated dielectric surface using nonrelativistic [18] and relativistic [29] approximations of the fluctuation-electromagnetic theory. It is easy to show that these results can be used to obtain the intensity of radiation generated by rotating particle. Let a particle is surrounded by a spherical surface $\sigma$ with sufficiently large radius, so that the electromagnetic field on $\sigma$ represents the wave field (Fig. 1). Using the law of energy conservation of electromagnetic field within the volume V restricted by $\sigma$ one can write [23]

$$\oint S_n \, d\sigma = -\int_V \langle \mathbf{j}\mathbf{E} \rangle d^3 r = I_1 - I_2 = I \tag{2}$$

where $\mathbf{S} = \dfrac{c}{4\pi}\langle \mathbf{E} \times \mathbf{H} \rangle$ is the Pointing vector of fluctuating electromagnetic field with components $\mathbf{E}$, $\mathbf{H}$, and $\mathbf{j}$ is the density current. The angular brackets in (2) denote the total quantum-statistical averaging, $I_1$ and $I_2$ denote the resultant intensities of emission and absorption of photons. Moreover, the volume integral in (2) can be evaluated over the volume of the particle, since the dielectric is assumed to be lossless. According to [9], for fluctuating dipole, the Joule dissipation integral in (2) can be rewritten in the form

$$\int_V \langle \mathbf{j} \cdot \mathbf{E} \rangle d^3 r = \langle \dot{\mathbf{d}} \cdot \mathbf{E} + \dot{\mathbf{m}} \cdot \mathbf{H} \rangle \equiv \frac{dQ}{dt} \tag{3}$$

where the points above $\mathbf{d}$ and $\mathbf{m}$ denote the time derivatives. Using (2) and (3) yields

$$I = I_1 - I_2 = -\frac{dQ}{dt} \ . \tag{4}$$

Equation (4) is a special case of the more general result [23]

$$I = -F_x V - dQ/dt , \tag{5}$$

obtained earlier for the radiation power from a particle moving in blackbody radiation. Here $F_x$ is the tangential force acting on the particle and $dQ/dt$ is determined from (3) under the condition of translational motion with velocity $V$. In the case without rotation, the condition $T_1 = T_2 = 0$ leads to $dQ/dt = 0$ and $I = -F_x V$ . It is the latter was used to calculate the intensity of Cherenkov's radiation in the case when the particle is moving near a transparent dielectric [24].

According to [29] and (4), when rotating particle is located at distance $z_0$ from the surface, $dQ/dt$ and $I$ are given by

$$\frac{dQ}{dt} = \frac{\hbar}{4\pi^2} \int_{-\infty}^{+\infty} d\omega\omega \int d^2k \sum_{s=e,m} \alpha_s''(\omega_+) \times$$

$$\times \mathrm{Im}\left[\frac{\exp(-2q_0 z_0)}{q_0} R_s(\omega,\mathbf{k})\right] \cdot \left[\coth\left(\frac{\hbar\omega}{2k_B T_2}\right) - \coth\left(\frac{\hbar\omega_+}{2k_B T_1}\right)\right] +$$

$$+ \frac{\hbar}{4\pi^2} \int_{-\infty}^{+\infty} d\omega\omega \int d^2k \sum_{s=e,m} \alpha_s''(\omega) \times$$

$$\times \mathrm{Im}\left[\frac{\exp(-2q_0 z_0)}{q_0} k^2 \Delta_s(\omega)\right] \cdot \left[\coth\left(\frac{\hbar\omega}{2k_B T_2}\right) - \coth\left(\frac{\hbar\omega}{2k_B T_1}\right)\right] \quad (6)$$

$$R_e(\omega,k) = \left(k^2 - \frac{\omega^2}{c^2}\right)\Delta_e + \frac{\omega^2}{c^2}\Delta_m, \quad R_m(\omega,k) = \left(k^2 - \frac{\omega^2}{c^2}\right)\Delta_m + \frac{\omega^2}{c^2}\Delta_e \quad (7)$$

$$\Delta_e = \frac{n^2 q_0 - q}{n^2 q_0 + q}, \quad \Delta_m = \frac{q_0 - q}{q_0 + q}, \quad q_0 = \sqrt{k^2 - \omega^2/c^2}, \quad q = \sqrt{k^2 - n^2\omega^2/c^2}, \quad \omega_+ = \omega + \Omega \quad (8)$$

In the case $\Omega = 0$ Eq. (6) describes the rate of particle heating (cooling) with allowance for both evanescent and propagating electromagnetic modes.

### 3. Frictional torque and energy balance

To clarify the issue of energy balance, we write down the Joule dissipation integral in frame of reference of rotating particle, $\Sigma'$:

$$\int \langle \mathbf{j}' \cdot \mathbf{E}' \rangle d^3 r' = \langle \dot{\mathbf{d}}' \cdot \mathbf{E}' + \dot{\mathbf{m}}' \cdot \mathbf{H}' \rangle \quad (9)$$

The integration is performed over the volume of the particle, and the result obviously determines the rate of particle heating, $dQ'/dt$ ($t \equiv t'$ (own time of particle) since the rotation is nonrelativistic). The values $\mathbf{d}, \mathbf{m}, \mathbf{E}, \mathbf{H}$ in (3) and their equivalents in (9) are related by the matrix of rotation around the $Z$ axis (Fig. 1) : $E_i = A_{ik} E_k'$, etc.,

$$A = \begin{pmatrix} \cos\Omega t & -\sin\Omega t & 0 \\ \sin\Omega t & \cos\Omega t & 0 \\ 0 & 0 & 1 \end{pmatrix} \quad (10)$$

Using (10) yields

$$\dot{\mathbf{d}}\mathbf{E} + \dot{\mathbf{m}}\mathbf{H} = \dot{\mathbf{d}}' \cdot \mathbf{E}' + \dot{\mathbf{m}}' \cdot \mathbf{H}' + (d_x' E_y' - d_y' E_x') + (m_x' H_y' - m_y' H_x') = \dot{\mathbf{d}}' \cdot \mathbf{E}' + \dot{\mathbf{m}}' \cdot \mathbf{H}' + M_z \cdot \Omega \quad (11)$$

In (12), $M_z = (\mathbf{d} \times \mathbf{E} + \mathbf{m} \times \mathbf{H})_z$ is the torque acting on the particle in frame $\Sigma$. With allowance for (3), (4), (9) and (11) one obtains

$$-M_z \cdot \Omega = dQ'/dt + I \tag{12}$$

From (13) it follows that the work of frictional torque is spent on the particle heating and radiation, as it should be.

According to [29], the torque $M_z$ has the form

$$M_z = -\frac{\hbar}{2\pi^2} \int_{-\infty}^{+\infty} d\omega \int d^2k \sum_{s=e,m} \alpha_s''(\omega_+) \mathrm{Im}\left[\frac{\exp(-2q_0 z_0)}{q_0} R_s\right] \cdot \left[\coth\left(\frac{\hbar\omega}{2k_B T_2}\right) - \coth\left(\frac{\hbar\omega_+}{2k_B T_1}\right)\right] \tag{13}$$

Equations (6), (12) and (13) provide a comprehensive description of radiation, friction and heating of rotating particle caused by fluctuation-electromagnetic interaction with transparent dielectric plate.

### 4. Zero temperature case

In the limiting case $T_1 = T_2 \to 0$, performing some transformations of the integrals in (6), one obtains

$$I^{(0)} = -\frac{\hbar}{\pi} \int_0^\Omega d\omega\, \omega \int_0^{\omega n/c} dk\, k \sum_{s=e,m} \alpha_s''(\Omega - \omega) \mathrm{Im}\left[\frac{\exp(-2q_0 z_0)}{q_0} R_s\right] \tag{14}$$

Analogously to that, in the limiting case $T_1 = T_2 \to 0$ Eq. (13) reduces to

$$M_z^{(0)} = \frac{2\hbar}{\pi} \int_0^\Omega d\omega \int_0^{\omega n/c} dk\, k \sum_{s=e,m} \alpha_s''(\Omega - \omega) \mathrm{Im}\left[\frac{\exp(-2q_0 z_0)}{q_0} R_s\right] \tag{15}$$

It is convenient to rewrite Eqs. (14) and (15) in the form

$$I^{(0)} = -\frac{\hbar}{\pi c^3} \int_0^\Omega d\omega\, \omega^4 \sum_{s=e,m} \alpha_s''(\Omega - \omega) \psi_s(n, \omega z_0/c) \tag{16}$$

$$M_z = \frac{2\hbar}{\pi c^3} \int_0^\Omega d\omega \omega^3 \sum_{s=e,m} \alpha_s''(\Omega - \omega)\psi_s(n, \omega z_0/c) \tag{17}$$

$$\psi_e(n,x) = \int_0^n dt\, t\, \text{Im}\left[\frac{\exp(-2x\sqrt{t^2-1})}{\sqrt{t^2-1}}\left((t^2-1)\frac{n^2\sqrt{t^2-1}-\sqrt{t^2-n^2}}{n^2\sqrt{t^2-1}+\sqrt{t^2-n^2}} + \frac{\sqrt{t^2-1}-\sqrt{t^2-n^2}}{\sqrt{t^2-1}+\sqrt{t^2-n^2}}\right)\right] \tag{18}$$

$$\psi_m(n,x) = \int_0^n dt\, t\, \text{Im}\left[\frac{\exp(-2x\sqrt{t^2-1})}{\sqrt{t^2-1}}\left((t^2-1)\frac{\sqrt{t^2-1}-\sqrt{t^2-n^2}}{\sqrt{t^2-1}+\sqrt{t^2-n^2}} + \frac{n^2\sqrt{t^2-1}-\sqrt{t^2-n^2}}{n^2\sqrt{t^2-1}+\sqrt{t^2-n^2}}\right)\right] \tag{19}$$

In many practical situations, namely at $\Omega < 10^9\ s^{-1}$ and $R(\mu m) << z_0(\mu m) << 1.5 \cdot 10^4/n(\mu m)$, the approximation $\exp(-2q_0 z_0) \cong 1$ can be used. In this case, the functions $\psi_{e,m}(n,x)$ are negative, being independent of $x$, while $I^{(0)}$ and $M_z$ are independent upon the distance $z_0$. Fig. 2 shows the functions $\psi_{e,m}(n,0)$. One can see that $|\psi_m(n,0)| >> |\psi_e(n,0)|$ at $n > 3$, i. e. the emission from magnetically polarized particle may be much more intense than emission from electrically polarized particle. Moreover, as compared to the radiation from a particle rotating in vacuum [13], the proper enhancement factor is $|\psi_{e,m}(n,0)| >> 1$ at $n > 10$.

The driving factor of radiation is the kinetic energy of rotation, but the rate of kinetic energy loss is not equal to $I^{(0)}$, since from (16), (17) it follows $-M_z\Omega > I^{(0)}$. According to (13), another part of the kinetic energy is converted into thermal (internal) energy, $dQ'/dt$ of the particle. Therefore, the state with $T_1 = 0$ is unstable.

## 5. Finite temperature case and numerical examples

We illustrate further these results by performing numerical calculations using a simple approximation for the particle polarizability in the low-frequency limit:

$$\alpha''(\omega) = R^3\, \text{Im}\left(\frac{\varepsilon(\omega)-1}{\varepsilon(\omega)+2}\right) \approx R^3 A\omega \tag{20}$$

For particles of sufficiently small size, such response represents the first term of the Mie series, and we can neglect radiation corrections in the effective particle polarizability. Eq. (20) can be used for metallic particles with $\varepsilon(\omega) = 1 + i4\pi\omega/\sigma_0$, $A = 3/4\pi\sigma_0$, where $\sigma_0$ is the static conductivity. It is also suitable for dielectric particles like $SiO_2$ at frequencies below the transverse phonon frequency. We adopt $\sigma_0 = 2.07 \cdot 10^{14}\ s^{-1}$ corresponding to graphite ($A = 1.15 \cdot 10^{-15}\ s$) and $A = 3.6 \cdot 10^{-15}\ s$ for $SiO_2$, using the analytical approximation for $\varepsilon(\omega)$ reported in [30].

Substituting (20) into (6), (13) at $x \to 0$ yields

$$I = -\frac{\hbar A R^3 \psi_e(n,0)}{2\pi c^3} \left[ \frac{\vartheta_1^6 - \vartheta_2^6}{126} + \frac{\Omega^2 \vartheta_1^4}{10} + \frac{\Omega^4 \vartheta_1^2}{6} + \frac{\Omega^6}{15} \right] \qquad (21)$$

$$M_z = \frac{\hbar A R^3 \psi_e(n,0)}{2\pi c^3} \left[ \frac{3\vartheta_1^4 + \vartheta_2^4}{30} \Omega + \frac{\vartheta_1^2}{3} \Omega^3 + \frac{\Omega^5}{5} \right] \qquad (22)$$

where $\vartheta_{1,2} = \frac{2\pi k_B T_{1,2}}{\hbar}$. Using (12), (21), (22) one obtains

$$\frac{dQ'}{dt} = -\frac{\hbar A R^3 \psi_e(n,0)}{2\pi c^3} \left[ \frac{\vartheta_2^6 - \vartheta_1^6}{126} + \frac{\Omega^2 \vartheta_2^4}{30} + \frac{\Omega^4 \vartheta_1^2}{6} + \frac{2\Omega^6}{15} \right] \qquad (23)$$

One can easily verify that Eqs. (21) and (22) agree with (16), (17) at $T_1 = T_2 = 0$ (with allowance for (20)).

Since the state $T_1 = 0$ is unstable (see Eq. (23)), the particle is heated and acquires the effective temperature $T_1 \approx 2.22\Omega$ corresponding to the condition $dQ'/dt = 0$, owing to the condition $\theta_2 = 0$ Then, according to Eq. (12), the frictional work is fully converted into radiation, $-M_z \Omega = I$. This is analogous to the decelerating process upon rotational motion in vacuum [13], and thermal emission from a particle moving in blackbody radiation [23]. When a particle rotates in cold vacuum ($\theta_2 = 0$), its effective temperature is $\theta_1 = 0.867\Omega$ [13], and the power radiated (in ours notations) is $I = 0.054 \hbar A R^3 \Omega^6 / c^3$. In the case under consideration, from (21) it follows $I = 0.683 \hbar A R^3 \Omega^6 |\psi_e(n,0)| / c^3$, i. e. the power radiated can be much greater and increases with increasing $n$ (see Fig. 2). According to [18, 23, 29], the characteristic time needed to reach the state of thermal equilibrium is always much less than the characteristic time of deceleration. Therefore, the particle dynamics and radiation can be analyzed at fixed values of temperatures $T_1$ and $T_2$ for a long period of time.

In the case $T_2 \neq 0$, the equilibrium particle temperature depends on $T_2$ and $\Omega$, whereas Eq. (21) determines the difference between the emitted and absorbed power of electromagnetic radiation (the terms depending on $\vartheta_1$ and $\vartheta_2$, respectively). Figs. 3, 4 show the values of $I$ calculated by Eq. (21) and normalized to the factor $I_0 = \frac{\hbar A R^3 |\psi_e(n,0)| \vartheta_2^6}{2\pi c^3}$, corresponding to the equilibrium temperature $T_1$ as a function of $\Omega/\vartheta_2$. Normalized particle temperature at equilibrium ($T_1/T_2$) is shown in the insets to Figs. 3, 4 as a function of $\Omega/\vartheta_2$. As we can see, in contrast to the case of particle rotation in vacuum [13], the equilibrium particle temperature increases with increasing $\Omega$ even if $\Omega < \vartheta_2$. The numerical values of $I_0$ for silicon dioxide particles with radii $50\,nm$ are listed in the Table at various $n$ and $T_2$.

Table

Intensity factor $I_0 \times 10^{26}$ $(W)$ for a spherical SiO$_2$ particle with a radius of $50\,nm$.

| $T_2, K$ \ $n$ | 3 | 10 | 30 | 50 |
|---|---|---|---|---|
| 0.1 | $1.3 \cdot 10^{-6}$ | $5.9 \cdot 10^{-6}$ | $1.9 \cdot 10^{-5}$ | $6.4 \cdot 10^{-5}$ |
| 1.0 | 1.3 | 5.9 | 19 | 64 |
| 10 | $1.3 \cdot 10^6$ | $5.9 \cdot 10^6$ | $1.9 \cdot 10^7$ | $6.4 \cdot 10^7$ |
| 100 | $1.3 \cdot 10^{12}$ | $5.9 \cdot 10^{12}$ | $1.9 \cdot 10^{13}$ | $6.4 \cdot 10^{13}$ |

For graphite particles with the same radius, the data of the table should be used with a factor of 0.32. For highly conductive metallic particles with radii larger than $10\,nm$, the contribution of the magnetic polarization becomes dominant. Therefore, with allowance for the fact that $|\psi_m(n,x)| \gg |\psi_e(n,x)|$, we can expect that intensity of radiation can be even higher than in the cases considered above.

The spectral distributions of radiated photons will be peaked at the positions of the maximal frequencies of radiation, namely, $2.22\Omega$ in the case $T_1 = T_2 = 0$ and the frequencies corresponding to the equilibrium temperature of particle, depending on $T_2$ and $\Omega$. We believe that these properties are challenging for experimental verification.

It is worth to discuss briefly the case of large distances $z_0$. It is interesting that the functions $\psi_{e,m}(n,x)$ exhibit an oscillating behavior (Fig. 5). Due to this some ranges of the frequency in the integral expressions (6), (17) can give a positive contribution to the torque $M_z$. Simultaneously with that, the analogous contrubutions to $I \equiv I_1 - I_2$ become negative. This means that a particle may speed up the rotation for some time, increasing the angular velocity and absorbing the energy from radiation background. This issue requires further deeper investigation.

**6. Conclusions**

We have calculated the intensities of radiation from particles rotating in close vicinity to a transparent dielectric plate within the framework of fluctuation-electromagnetic theory. The characteristic frequencies of radiation are tuned to the frequencies of rotation or depend on the temperature of the plate and rotation frequency, as long as the particle reaches the state of quasiequilibrium with the proper effective temperature. In this case the particle experiences the action of braking torque and the kinetic energy of rotation is fully converted into radiation. These results are challenging for experimental verification in future.

**References**

[1] V. G. Polevoi, Sov. Phys. JETP 71(6), 1119 (1990).

[2] J. B. Pendry, J. Phys.: Condens. Matter 11, 345 (1997); J. Mod. Opt. 45, 2389 (1998).

[3] M. S. Tomassone, A. Widom, Phys. Rev. B56, 4938 (1997).

[4] M. Kardar and R. Golestanian, Rev. Mod. Phys. 71(4), 1233 (1999).

[5] A. I. Volokitin, B.N. J. Persson, J. Phys.: Condens. Matter 11, 345 (1999).

[6] A. A. Kyasov, G. V. Dedkov, Nucl. Instr. Meth. B195, 247 (2002); Phys. Low-Dim. Struct.1/2, 1 (2003).

[7] P. C. W. Davies, J. Opt. B.: Quant. Semiclass. Opt. 7(3): S40 (2005).

[8] A. I. Volokitin, B. N. J. Persson, Rev. Mod. Phys. 79, 1291 (2007).

[9] G. V. Dedkov, A. A. Kyasov, J. Phys.: Condens. Matter 20, 354006 (2008).

[10] T. G. Philbin, U. Leonhardt, New J. Phys. 11, 033035 (2009).

[11] J. B. Pendry, New J. Phys., 12, 033028 (2010).

[12] G. Barton, New J. Phys. 12, 113045 (2010).

[13] A. Manjavacas and F. G. Garcia de Abajo, Phys. Rev. Lett. 105, 113601 (2010); Phys. Rev. A82 063827 (2010).

[14] F. Intravaia, C. Henkel, and M. Antezza, in: Casimir Physics (Lecture Notes in Physics, 834 (2011) 345), ed. D.A.R. Dalvit, P.W. Milonni, D. Roberts and F. Da Rosa (Berlin: Springer), Ch. 11, pp. 345-391.

[15] J. S. Høye and I. Brevik, Eur. Phys. J. D66, 149 (2012); Entropy 15, 3045 (2013); J. Phys.: Condens. Matter 27, 214008 (2015).

[16] G. Pieplow and C. Henkel, New J. Phys., 15, 023027 (2013).

[17] R. Zhao, A. Manjavacas, F. J. Garcia de Abajo, and J. B. Pendry, Phys. Rev. Lett. 109, 123604 (2012).

[18] G. V. Dedkov, A. A. Kyasov, Europhys. Lett. 99, 6402 (2012).

[19] X. Chen, Int. J. Mod. Phys. B27(14), 1350066 (2013); ibid.: B28(15), 1492002 (2014).

[20] M. F. Maghrebi, R. L. Jaffe, and M. Kardar, Phys. Rev. Lett. 108, 230403 (2012)

[21] M. F. Maghrebi, R. Golestanian, M. Kardar, Phys. Rev. A88, 042509 (2013); A90, 012515 (2014).

[22] K. A. Milton, J. S. Høye and I. Brevik, Symmetry 8, 29 (2016).

[23] G. V. Dedkov, A. A. Kyasov, Phys. Scripta 89(10), 105501 (2014).

[24] G. Pieplow and C. Henkel, J. Phys.: Condens. Matter 27, 214001 (2015).

[25] A.I. Volokitin, Phys. Rev. A91, 032505 (2015); arXiv: 1511.06573.

[26] G. V. Dedkov, A. A. Kyasov, Int. J. Mod. Phys. B29(32) 1550237 (2015).

[27] G.V. Dedkov, A. A. Kyasov, Tech. Phys. Lett. 42(1), 8 (2016); arXiv: 1504.01588.

[28] B. E. Kane, Phys. Rev. B82, 115441 (2010).

[29] A. A. Kyasov, G. V. Dedkov, arXiv: 1303.7421.

[30] Chen D.-Z. A., Hamam R., Soljacic M., Joan Nopoulos J. D. and Chen Gang, Appl. Phys. Lett., 90, 181921 (2007).

Fig. 1.

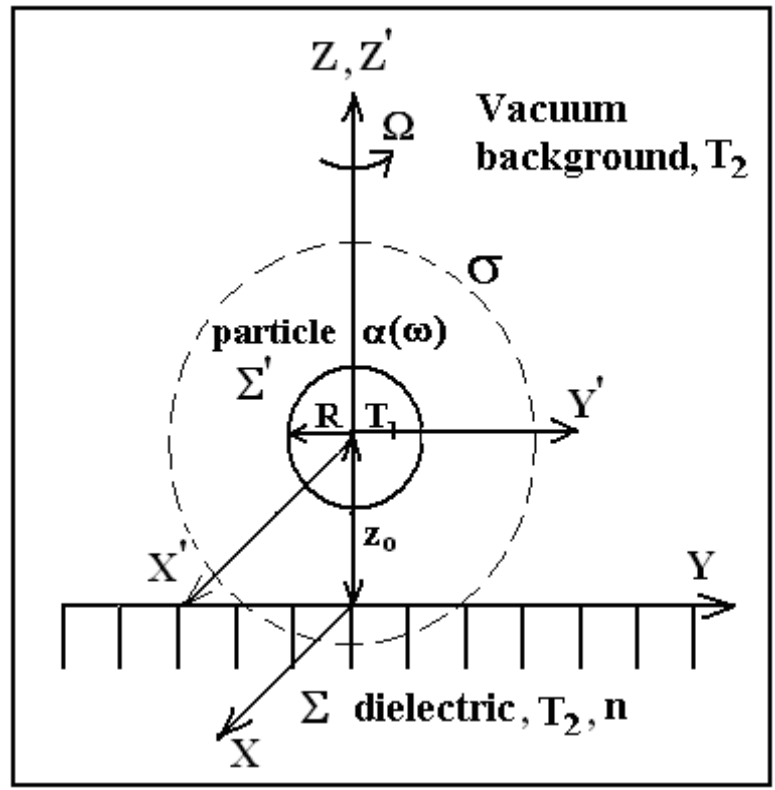

Fig. 2.

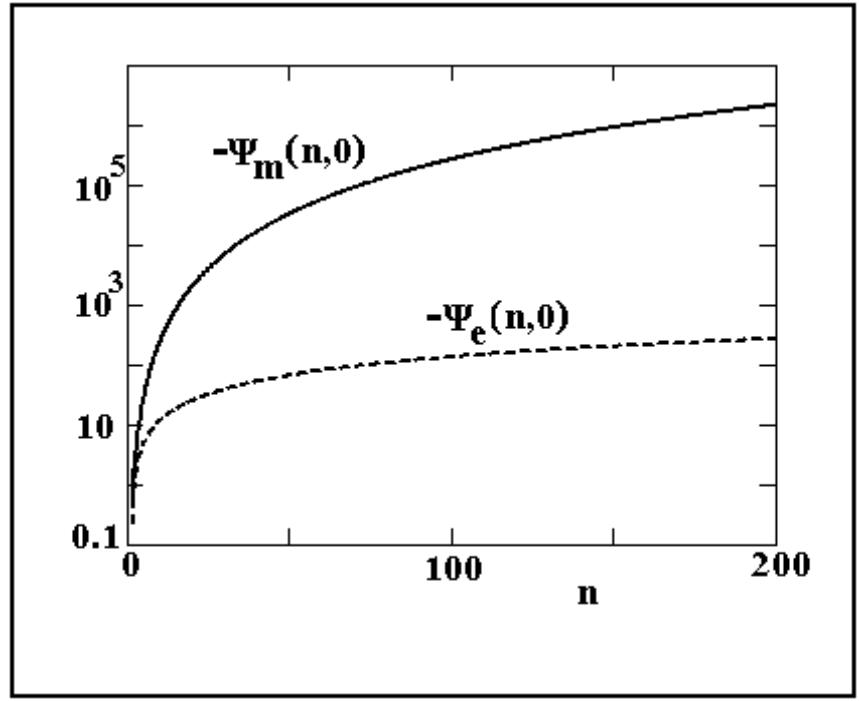

Fig. 3

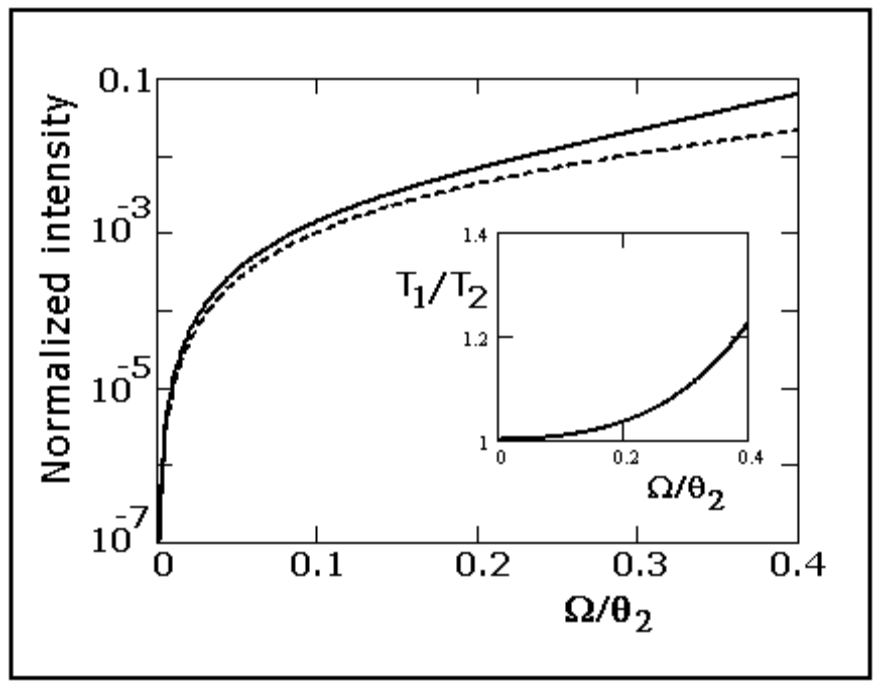

Fig. 4

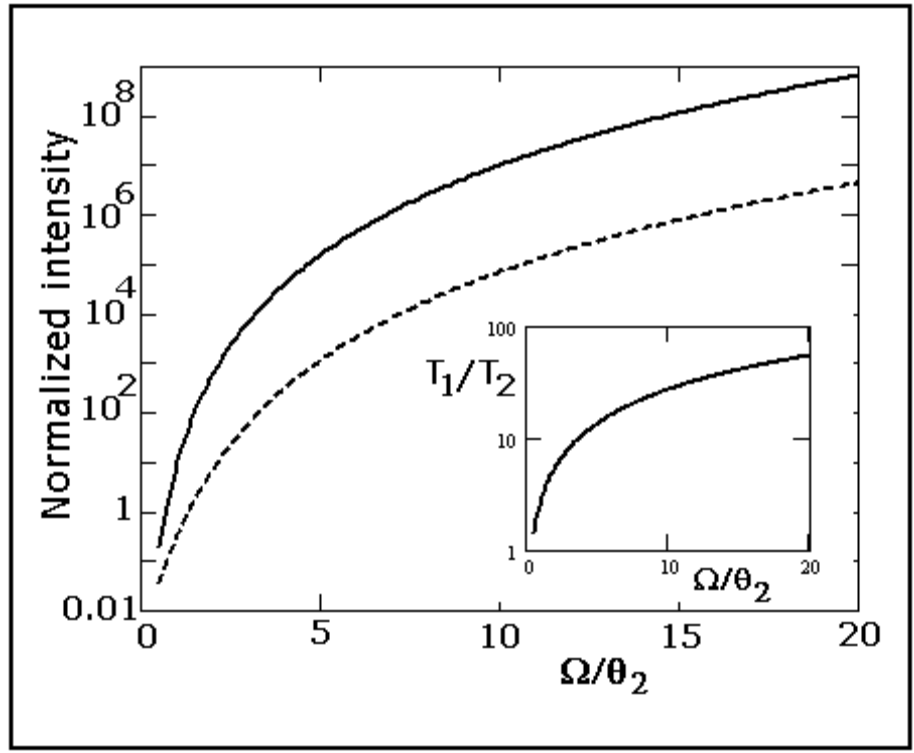

Fig.5

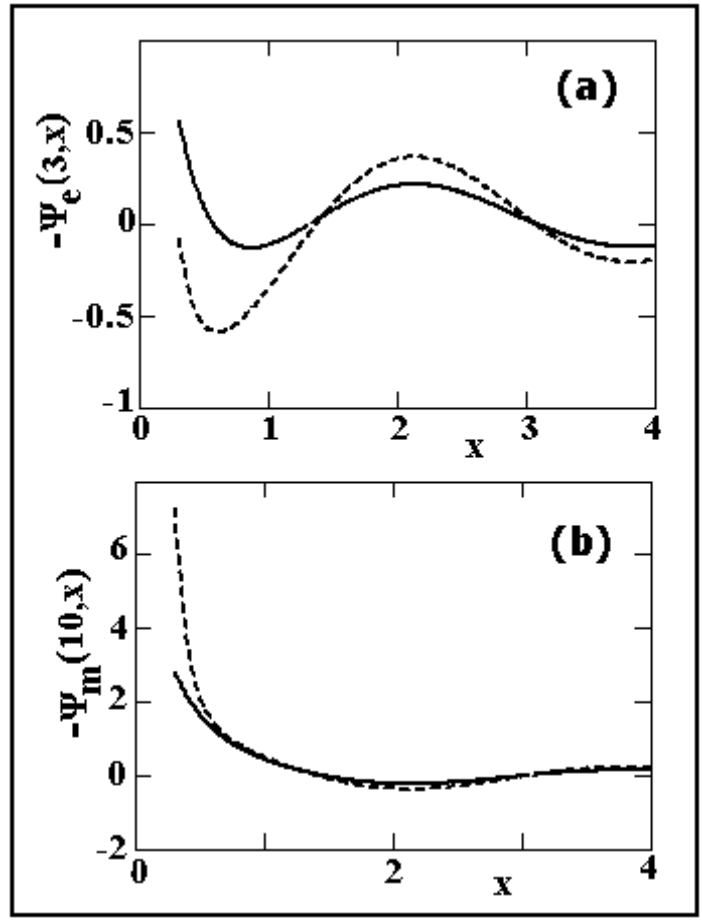

FIGURE CAPTIONS

Fig.1. Geometrical configuration and the frames of reference $\Sigma, \Sigma'$.

Fig. 2. Functions $-\psi_e(n,x)$ (solid line) and $-\psi_m(n,x)$ (dashed line) at $x=0$.

Fig.3. Normalized intensity $I/I_0$ and equilibrium particle temperature $T_1/T_2$ as functions of $\Omega/\vartheta_2$. Solid lines correspond to the equilibrium particle temperature shown in the insets, dashed lines correspond to the state $T_1 = T_2$. The intensity factor is $I_0 = \dfrac{\hbar A R^3 |\psi_e(n,0)| \vartheta_2^6}{2\pi c^3}$.

Fig. 4. Same as in Fig. 3.

Fig. 5. Functions $-\psi_e(n,x)$ and $-\psi_m(n,x)$ at various $x$. Solid and dashed curves correspond to and $n=3$ and $n=10$.